\documentclass[aps,prb,twocolumn,superscriptaddress,floatfix]{revtex4-2}
\usepackage[utf8]{inputenc}
\usepackage[T1]{fontenc}
\usepackage{amsmath,amssymb,graphicx,booktabs}
\usepackage[colorlinks=true,linkcolor=blue,urlcolor=blue,citecolor=blue]{hyperref}

\begin{document}

\title{Learning to control and predict light-driven electron--phonon dynamics}

\author{Matteo Puviani}
\thanks{Present address: Quantinuum Inc., London, United Kingdom. This work was carried out during the author's affiliation with the Max Planck Institute for the Science of Light in 2024-2025, including a guest appointment in 2025.}
\affiliation{Max Planck Institute for the Science of Light, 91058 Erlangen, Germany}

\begin{abstract}
Driving lattice vibrations provides a direct route to manipulating electronic correlations through nonlinear electron--phonon coupling. Finding suitable driving fields, however, requires control over the coupled dynamics beyond a simple choice of pulse amplitude and frequency. Here we develop a differentiable framework that combines field optimization with neural-network prediction in a two-site electron--phonon model. We compare piecewise-constant, B-spline, feed-forward, and recurrent control parameterizations, optimized to increase the accumulated electronic double occupation. For the dissipation-free model at zero Hubbard interaction, all the methods increase double occupation, with the B-spline raising the mean total double occupation of approximately 30\%. Alongside control optimization, a separate recurrent network learns to predict the next-step double occupation and phonon observables from their preceding values and the applied field. These results establish a computational proof of concept for shaping local electronic correlations through a phonon drive and identify a promising route toward learned models for nonequilibrium materials control.
\end{abstract}

\maketitle

\section{Introduction}

Light-driven materials offer the possibility to manipulate electronic states on the time scale of their microscopic dynamics. Periodic fields can modify band structure and interactions through Floquet engineering \cite{Tsuji2008,Oka2009,GomezLeon2013,Puviani2016periodically,Castro2022}, while ultrashort pulses provide access to transient states that are difficult to reach in equilibrium. Examples include light-driven metal--insulator transitions \cite{Esin2020,Zahradnik2022}, topological states \cite{Dehghani2021}, and evidence for photoinduced superconductivity \cite{Budden2021}. Cavity-mediated control offers a complementary approach to modifying material properties through their electromagnetic environment \cite{Latini2021,Schlawin2022,bostroem2022controlling,Rokaj2022}.

Of particular interest is the selective excitation of lattice vibrations. When the phonon coordinate couples nonlinearly to the electronic density, driving the lattice can change the electronic response and favor configurations with enhanced double occupation \cite{Sentef2017Light}. Related nonlinear couplings also provide a microscopic route to controlling intertwined electronic and structural order \cite{Puviani2018Quantum}. The resulting dynamics is not generally captured by the instantaneous field strength alone: electronic correlations and lattice motion evolve together, making the shape and timing of the drive relevant control parameters. This raises a practical question: namely, on how it is possible to discover driving fields that exploit this coupled evolution to enhance a chosen electronic observable.

Optimal control addresses this question by turning pulse design into an optimization problem \cite{Khaneja2005}. Automatic differentiation makes it possible to evaluate control gradients through a numerical simulation and to optimize both directly parameterized fields and neural controllers \cite{Leung2017,Coopmans2021Majorana,Porotti2023}. Machine learning has further expanded the range of available strategies. Reinforcement learning has been applied to many-body state preparation, quantum feedback, and gate design \cite{Bukov2018RL,Foesel2018QuantumFeedback,Niu2019UniversalQuantumControl}; recurrent networks can exploit measurement histories for control \cite{Puviani2023boosting} or infer quantum dynamics from observations \cite{Flurin2020RNN}. These developments motivate the use of learned, history-dependent maps for controlling and representing coupled electron--phonon dynamics.

\begin{figure*}[t]
\centering
\includegraphics[width=\textwidth]{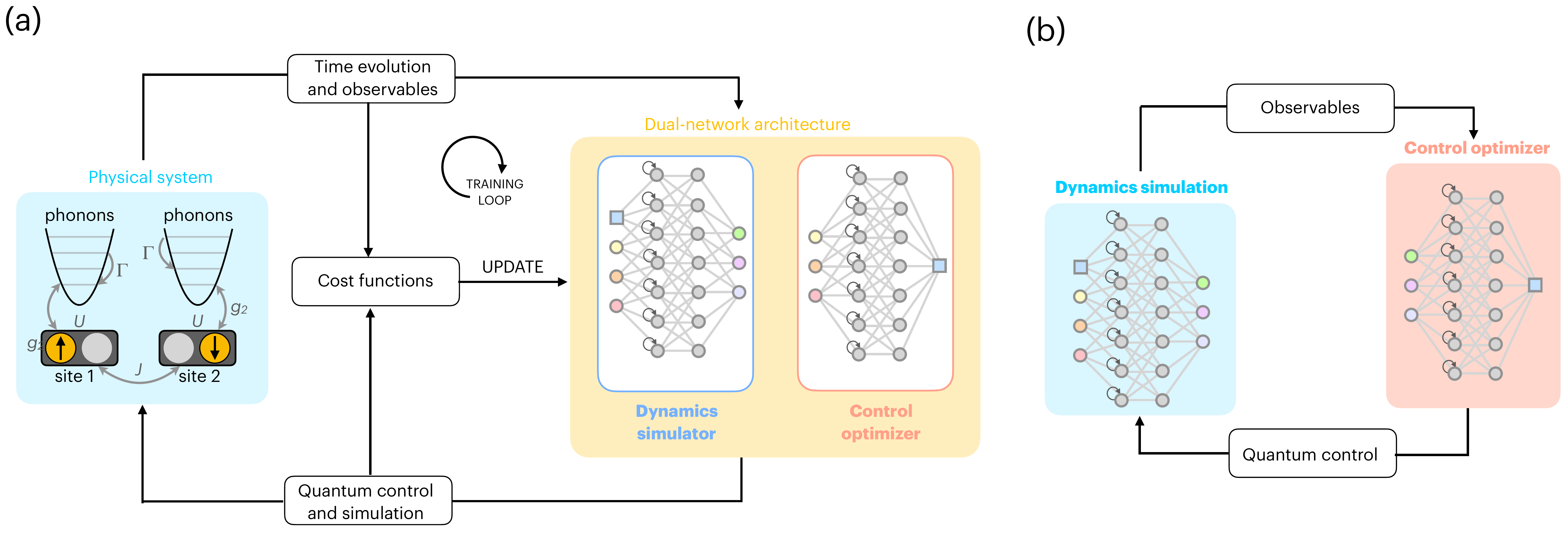}
\caption{Framework for differentiable control and learned dynamics. (a) During training, a quantum simulator supplies the observables and their response to the driving field. The control network is optimized by differentiating the accumulated double occupation through this simulated trajectory. A separate dynamics network receives the current observables and field and is trained against the next-step simulator outputs. The two networks use distinct objectives and optimizers. (b) Prospective deployment in which a learned dynamics model and controller are connected recursively.}
\label{scheme}
\end{figure*}

Here we study this possibility in a minimal two-site model with quadratic electron--phonon coupling. We optimize four representations of the phonon drive: a vector of independent amplitudes, a B-spline expansion, a feed-forward neural network (FNN), and a recurrent neural network (RNN). All four use the same double-occupation objective and gradients obtained through the physical simulator. A separate recurrent network is trained to predict the next-step observables, providing the second component of the architecture in Fig.~\ref{scheme}. The controller therefore learns how to drive the system, whereas the predictor learns how the selected observables respond.

The optimized trajectories show that pulse shaping can substantially increase the time-averaged double occupation relative to the chosen monochromatic reference. The recurrent controller is also evaluated beyond its training interval, while the dynamics network reproduces the main features of a sequence of one-step predictions. The two-site setting allows these ideas to be examined with full quantum dynamics in a truncated Hilbert space. It probes local electronic pairing tendencies and provides a useful starting point for extending learned control to larger and more realistic systems.

\section{Model and observables}

We consider two electronic sites at half filling, with one spin-up and one spin-down electron. Each site is coupled to a local phonon mode through the square of its displacement. The equilibrium Hamiltonian is given by \cite{Sentef2017Light}
\begin{align}
\hat H_0={}&-J\sum_{\sigma}\left(\hat c^\dagger_{1\sigma}\hat c_{2\sigma}+\mathrm{H.c.}\right)
+U\sum_{i=1}^{2}\hat n^{\mathrm{el}}_{i\uparrow}\hat n^{\mathrm{el}}_{i\downarrow}\nonumber\\
&+\Omega\sum_{i=1}^{2}\hat a_i^\dagger\hat a_i
+g_2\sum_{i=1}^{2}\hat n^{\mathrm{el}}_i\hat X_i^2,
\label{Ham0}
\end{align}
where $\hat n^{\mathrm{el}}_i=\sum_\sigma\hat c^\dagger_{i\sigma}\hat c_{i\sigma}$ and $\hat X_i=\hat a_i+\hat a_i^\dagger$. The operators $\hat c_{i\sigma}$ and $\hat a_i$ annihilate an electron and a phonon, respectively; $J$ is the electronic hopping, $U$ the on-site Hubbard interaction, $\Omega$ the bare phonon frequency, and $g_2$ the quadratic coupling. We set $\hbar=1$. For an inversion-odd phonon coordinate, an electronic-density coupling linear in $\hat X_i$ is forbidden by inversion symmetry, whereas the quadratic term is allowed \cite{Puviani2018Quantum}. An external field drives the two phonon modes equally,
\begin{equation}
\hat H(t)=\hat H_0+\alpha(t)\sum_{i=1}^{2}\hat X_i.
\label{Ham1}
\end{equation}
Here $\alpha(t)$ is the effective drive amplitude, with units of energy in this normalization; conversion to an electric field requires the mode's effective charge and displacement scale.

The quadratic coupling makes the local phonon potential depend on electronic occupation. Conversely, a driven phonon changes the dynamics of the electrons. This mutual dependence is the physical basis for controlling electronic observables through the lattice. Even at $U=0$, the coupled electron-phonon system is interacting via exchange of phonons.

The formulation also allows local phonon loss, described by
\begin{align}
\dot{\hat\rho}&=-i[\hat H(t),\hat\rho]+\Gamma\sum_{i=1}^{2}\mathcal D[\hat a_i]\hat\rho,\label{Lindblad}\\
\mathcal D[\hat a]\hat\rho&=\hat a\hat\rho\hat a^\dagger-\tfrac12\{\hat a^\dagger\hat a,\hat\rho\}.
\end{align}
All results below use $U=\Gamma=0$, $J=0.15$, $\Omega=0.5$, and $g_2=-0.05$. Each phonon mode is truncated to $N_{\mathrm{ph}}=20$ Fock states, giving a Hilbert-space dimension $4N_{\mathrm{ph}}^2=1600$ in the specified electronic sector. The field is held constant within time intervals of duration $\delta t=0.1$. In the absence of dissipation, a pure state can be propagated with the Schr\"odinger equation; the implementation for nonzero $\Gamma$ uses fourth-order Runge--Kutta integration of Eq.~\eqref{Lindblad}.

Our electronic control observable is the total double occupation,
\begin{equation}
\hat D=\sum_{i=1}^{2}\hat n^{\mathrm{el}}_{i\uparrow}\hat n^{\mathrm{el}}_{i\downarrow},
\qquad D(t)=\mathrm{Tr}[\hat\rho(t)\hat D].
\label{doublon}
\end{equation}
In the sector with one electron of each spin, $0\leq D\leq1$: it is the probability that both electrons occupy the same site, summed over the two possible sites. The site-averaged double occupation is $D/2$, and increasing $D$ thus favors local electronic pairing, which can lead to phase coherence or superconductivity.

The neural networks use the observable vector
\begin{equation}
\mathbf o_n=(D_n,n_n,X_n),\quad
n_n=\langle\hat a_1^\dagger\hat a_1\rangle_{t_n},\quad
X_n=\langle\hat X_1\rangle_{t_n}.
\label{observables}
\end{equation}
For the site-symmetric initial condition and symmetric drive considered here, the two sites have identical local phonon expectation values, so site 1 suffices. 

\section{Differentiable control}

\subsection{Objective and optimization}

Let $t_n=n \ \delta t$, with $M$ control intervals and total duration $T=M \ \delta t$. At time $t_n$, the amplitude $\alpha_n$ is selected and applied on $[t_n,t_{n+1})$. The simulator then provides the state and observables at $t_{n+1}$. We maximize the reward function 
\begin{equation}
\mathcal R(\theta)=\sum_{n=1}^{M}D_n,
\qquad \overline D=\frac{1}{M}\sum_{n=1}^{M}D_n,
\label{reward}
\end{equation}
where $\theta$ denotes the parameters of the chosen controller, and $D_n$ is the double occupation at time step $t_n$. On a fixed grid, $\delta t\mathcal R$ approximates $\int_0^T D(t)\,dt$ and $\overline D$ approximates its time average. These objectives have the same maximizers for fixed $M$ and $\delta t$; their gradient magnitudes differ by a constant factor.

Automatic differentiation propagates derivatives through the discretized dynamics and, where present, through the controller at every time step \cite{Leung2017,Porotti2023}. Adam \cite{Kingma2015Adam} is used to update the control parameters. This is a model-based optimization: it uses the response of the simulated quantum system and requires neither examples of optimal fields nor a learned reward estimator. The directly parameterized field is closely related to gradient-based pulse engineering \cite{Khaneja2005}, with accumulated double occupation replacing the more usual terminal-state or gate objective.

\subsection{Independent field amplitudes}

The most direct representation assigns one parameter to every control interval,
\begin{equation}
\alpha_n=F_n,\qquad \theta=(F_0,\ldots,F_{M-1}).
\end{equation}
This piecewise-constant vector provides considerable freedom to shape the drive, at the cost of a parameter count that grows with the control duration. The resulting protocol is defined on its optimized interval. Extending it would require an additional prescription, such as reoptimization or repetition.

\subsection{B-spline representation}

\begin{figure}[t]
\centering
\includegraphics[width=\columnwidth]{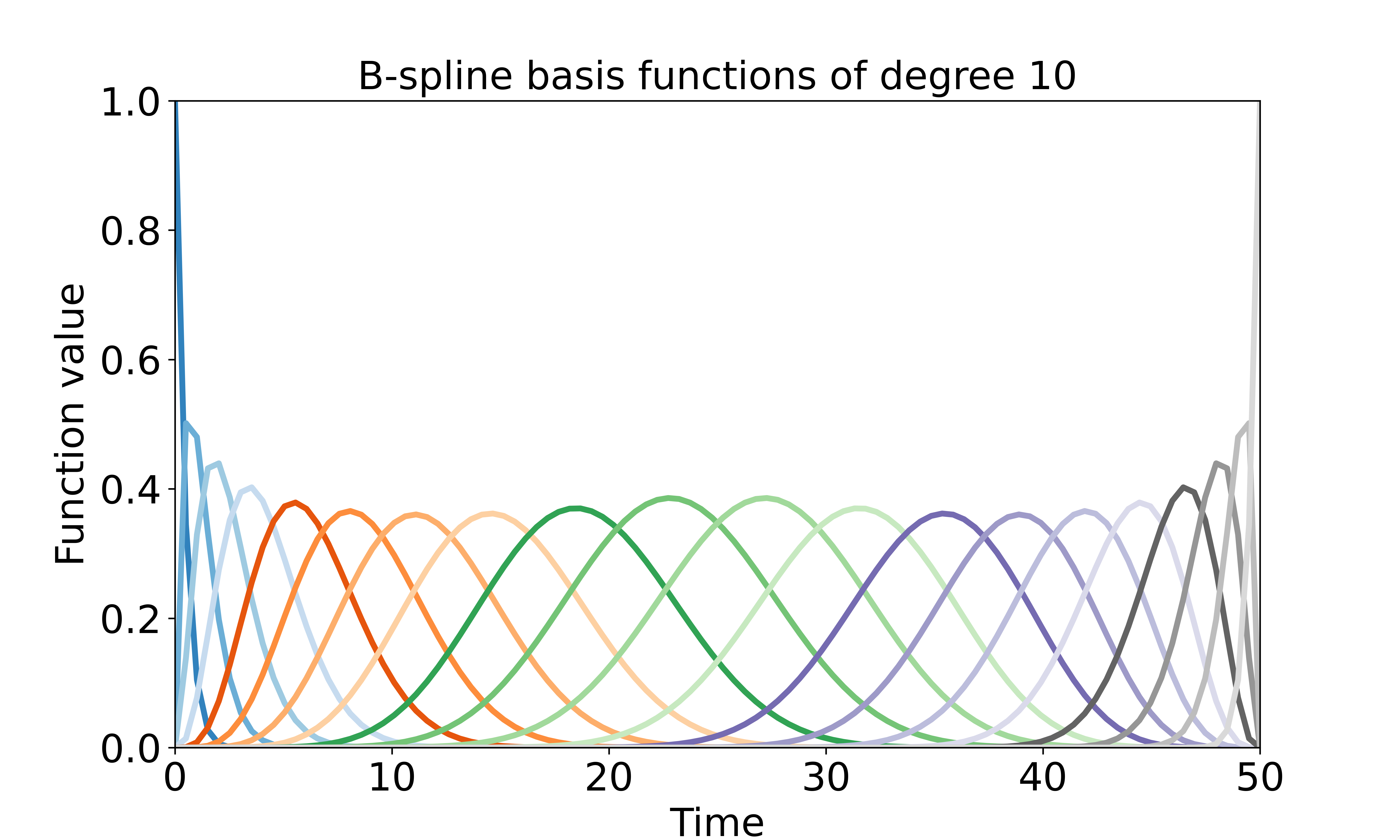}
\caption{The 20 degree-10 B-spline basis functions used to parameterize the drive, illustrated over $0\leq t\leq50$. The clamped knot vector contains 31 entries, including repeated endpoints. Optimizing their coefficients changes the field while preserving the structure of the basis.}
\label{Bspline}
\end{figure}

A more compact representation expands the drive in a smooth basis,
\begin{equation}
\alpha(t)=\sum_{q=0}^{19}c_qB_{q,10}(t),\qquad \alpha_n=\alpha(t_n).
\label{spline}
\end{equation}
We use 20 B-splines of degree 10, defined by a clamped knot vector with 31 entries: each endpoint has multiplicity 11, and the nine interior knots are uniformly spaced in $(0,T)$ (Fig.~\ref{Bspline}) \cite{deBoor2001}. Only the 20 coefficients $c_q$ are optimized. This representation couples neighboring field values and favors smoothly varying profiles, while retaining the common piecewise-constant propagation grid. The spline degree controls polynomial smoothness, but neither a spline basis nor a small parameter count alone imposes an experimental amplitude or bandwidth constraint.

\section{Neural control and prediction}

\subsection{Feed-forward controller}

A feed-forward network represents a nonlinear map from the current observables to the applied amplitude,
\begin{equation}
\alpha_n=f_\theta(\mathbf o_n).
\label{fnn}
\end{equation}
Its hidden layers successively transform an input vector according to $\mathbf z^{(\ell+1)}=\tanh(W^{(\ell)}\mathbf z^{(\ell)}+\mathbf b^{(\ell)})$, starting from $\mathbf z^{(0)}=\mathbf o_n$. The trainable weights and biases determine how combinations of observables are converted into a control signal. Nonlinear activations allow the network to represent responses beyond a linear feedback law, and backpropagation supplies derivatives with respect to these parameters \cite{Rumelhart1986}. The controller used here has two hidden layers with 30 hyperbolic-tangent units each and a scalar field output.

The FNN has no internal state carried between time steps: identical input vectors produce identical amplitudes for fixed parameters. Nevertheless, it generates a time-dependent field because the observables evolve. 

\subsection{Recurrent controller}

The selected observables need not uniquely specify the full quantum state. Two states can have the same $(D_n,n_n,X_n)$ but different correlations and hence different future responses. A recurrent controller can use temporal information to distinguish situations that an instantaneous map cannot resolve. It updates a hidden state and generates the field according to
\begin{align}
\mathbf h_n&=\mathrm{GRU}_{\theta}(\mathbf o_n,\mathbf h_{n-1}),\nonumber\\
\alpha_n&=g_\theta(\mathbf h_n).
\label{rnn}
\end{align}
The gated recurrent unit (GRU) uses learned update and reset gates to regulate the retention and replacement of information \cite{Cho2014GRU}. Its hidden state is a finite-dimensional summary of the input history, rather than a complete representation of the many-body wave function: this provides a useful inductive bias for a partially observed dynamical system.

The recurrent controller uses a 20-unit GRU, followed by two 30-unit hyperbolic-tangent layers and a single hyperbolic-tangent output. The final activation bounds the output in its network units. Its parameters are optimized by differentiating the same trajectory objective as for the other controls, including the dependence of later hidden states and fields on earlier inputs. Both the FNN and RNN can be iterated beyond their training horizon; the additional capability of the RNN is its learned memory. In the present comparison, only the RNN is shown over an extended interval.

\subsection{Learning a dynamics model}

The dual-network architecture pairs the recurrent controller with a separate recurrent predictor. At each step, the predictor receives $\mathbf o_n$ and $\alpha_n$, updates its own hidden state $\mathbf s_n$, and estimates $\mathbf o_{n+1}$:
\begin{equation}
(\widetilde{\mathbf o}_{n+1},\mathbf s_n)
=P_\phi(\mathbf o_n,\alpha_n,\mathbf s_{n-1}).
\label{predictor}
\end{equation}
The controller and predictor have independent parameter sets, $\theta$ and $\phi$, and separate Adam optimizers. During training, the simulator supplies the supervised targets for the predictor, while the controller is optimized through the physical trajectory.

The prediction objective is the weighted absolute-error loss
\begin{align}
\mathcal L_{\mathrm{dyn}}=\sum_{n=0}^{M-1}\big[&10|\widetilde D_{n+1}-D_{n+1}|\nonumber\\
&+|\widetilde n_{n+1}-n_{n+1}|\nonumber\\
&+|\widetilde X_{n+1}-X_{n+1}|\big].
\label{loss}
\end{align}
The factor of 10 increases the importance assigned to double occupation relative to the two phonon observables. We therefore evaluate the error of each observable separately. The reported predictor uses 200-step training episodes and a learning rate of $5\times10^{-6}$.

The assessment below uses \emph{teacher forcing}: at every step the predictor receives the preceding simulator observables. This tests conditional one-step prediction over an extended sequence. A free-running rollout would instead feed predicted observables back into Eq.~\eqref{predictor}; coupling that rollout to the controller would realize the loop in Fig.~\ref{scheme}(b).

\section{Results}

\subsection{Enhancing electronic double occupation}

\begin{figure}[t]
\centering
\includegraphics[width=\columnwidth]{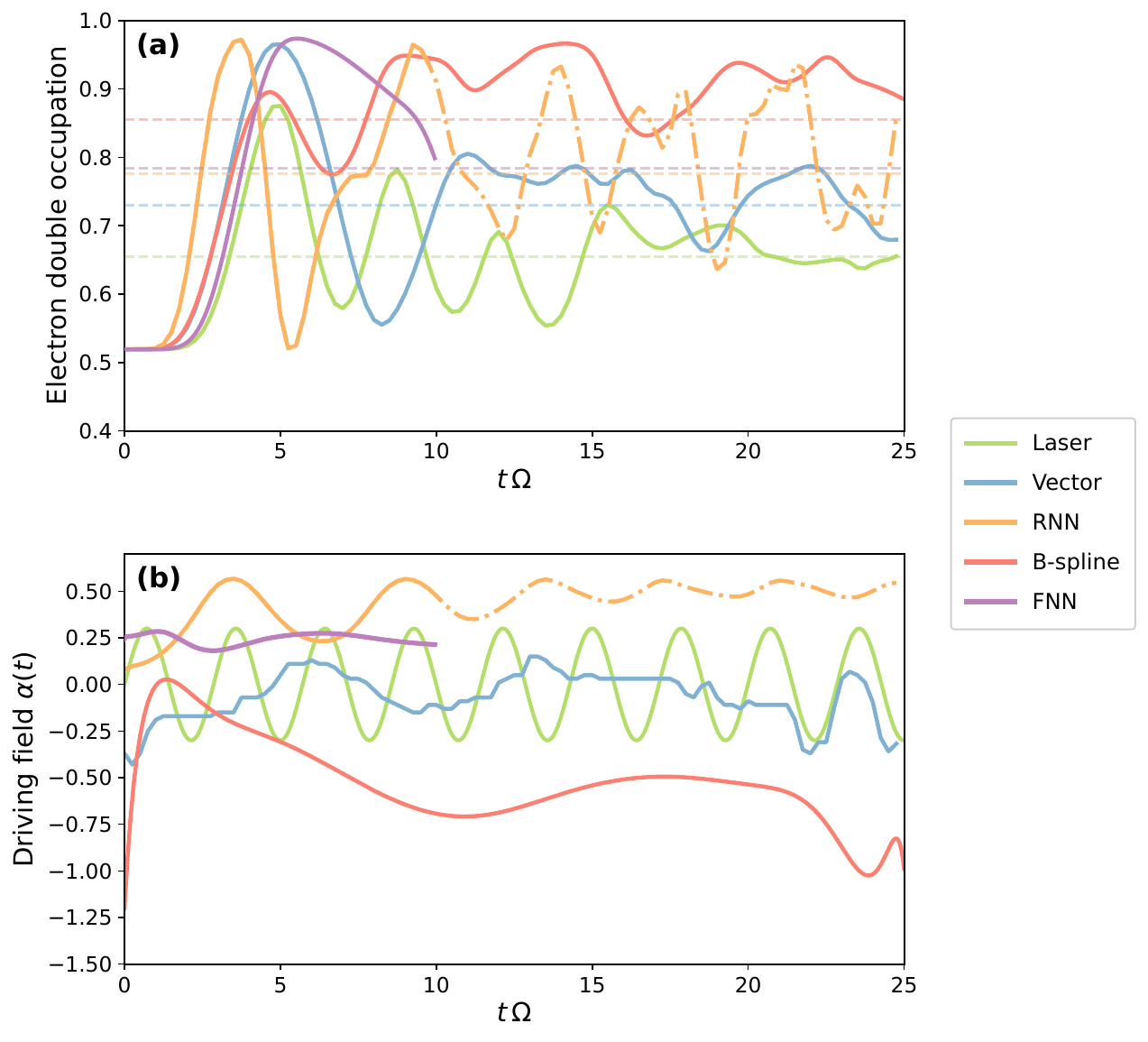}
\caption{Control of the two-site electron--phonon model at $U=\Gamma=0$. (a) Total double occupation for the monochromatic reference and four control parameterizations. Horizontal dashed lines indicate the reported means: 0.655 (laser), 0.729 (vector), 0.855 (B-spline), 0.776 (RNN), and 0.784 (FNN). The FNN trace covers only $0\leq t\Omega\leq10$; the other traces extend to $t\Omega=25$. Its mean therefore refers to a different interval. (b) Corresponding driving fields. Both neural controllers are trained to $t\Omega=10$; the dash-dotted RNN continuation shows evaluation beyond that horizon. The FNN is not evaluated over the longer interval in this figure.}
\label{Result1}
\end{figure}

Figure~\ref{Result1} compares representative optimized controls for the dissipation-free model with zero Hubbard interaction. The monochromatic reference produces an initial increase of double occupation followed by pronounced oscillations. Optimizing the field changes both the transient response and the subsequent evolution. In particular, the B-spline trajectory maintains an elevated double occupation over much of the displayed interval, showing how a compact, structured field can exploit the coupled electron--phonon dynamics.

\begin{table}[b]
\caption{Reported total-double-occupation averages for Fig.~\ref{Result1}. The changes are differences of the displayed rounded means. The FNN is listed separately because its averaging interval is shorter. These are individual trajectory results, without uncertainty estimates across training runs.}
\label{tab:control}
\begin{ruledtabular}
\begin{tabular}{lccc}
Control & Interval in $t\Omega$ & $\overline D$ & Change from laser\\
\hline
Monochromatic & $[0,25]$ & 0.655 & ---\\
Vector & $[0,25]$ & 0.729 & $+0.074$\\
B-spline & $[0,25]$ & 0.855 & $+0.200$\\
RNN & $[0,25]$ & 0.776 & $+0.121$\\
\hline
FNN & $[0,10]$ & 0.784 & --- \\
\end{tabular}
\end{ruledtabular}
\end{table}

Table~\ref{tab:control} summarizes the reported averages. Over the common displayed interval, the vector, B-spline, and recurrent controls increase $\overline D$ by 0.074, 0.200, and 0.121 relative to the laser reference. The largest increase in these trajectories is obtained with the B-spline representation, approximately 30\% relative to the reference value. This is a substantial change in the probability of finding the two electrons on the same site. The shorter FNN trajectory has mean 0.784.

The fields in Fig.~\ref{Result1}(b) differ substantially in amplitude and temporal structure. In particular, the larger excursions of the B-spline field may contribute to its higher double occupation. The present result therefore demonstrates the effectiveness of the selected shaped fields relative to the chosen monochromatic drive. A further comparison at fixed peak amplitude, integrated squared drive, bandwidth, and optimization budget would isolate the contribution of the representation itself. Such constraints can be included directly in the parameterization or through additional objective terms.

\begin{figure}[t]
\centering
\includegraphics[width=\columnwidth]{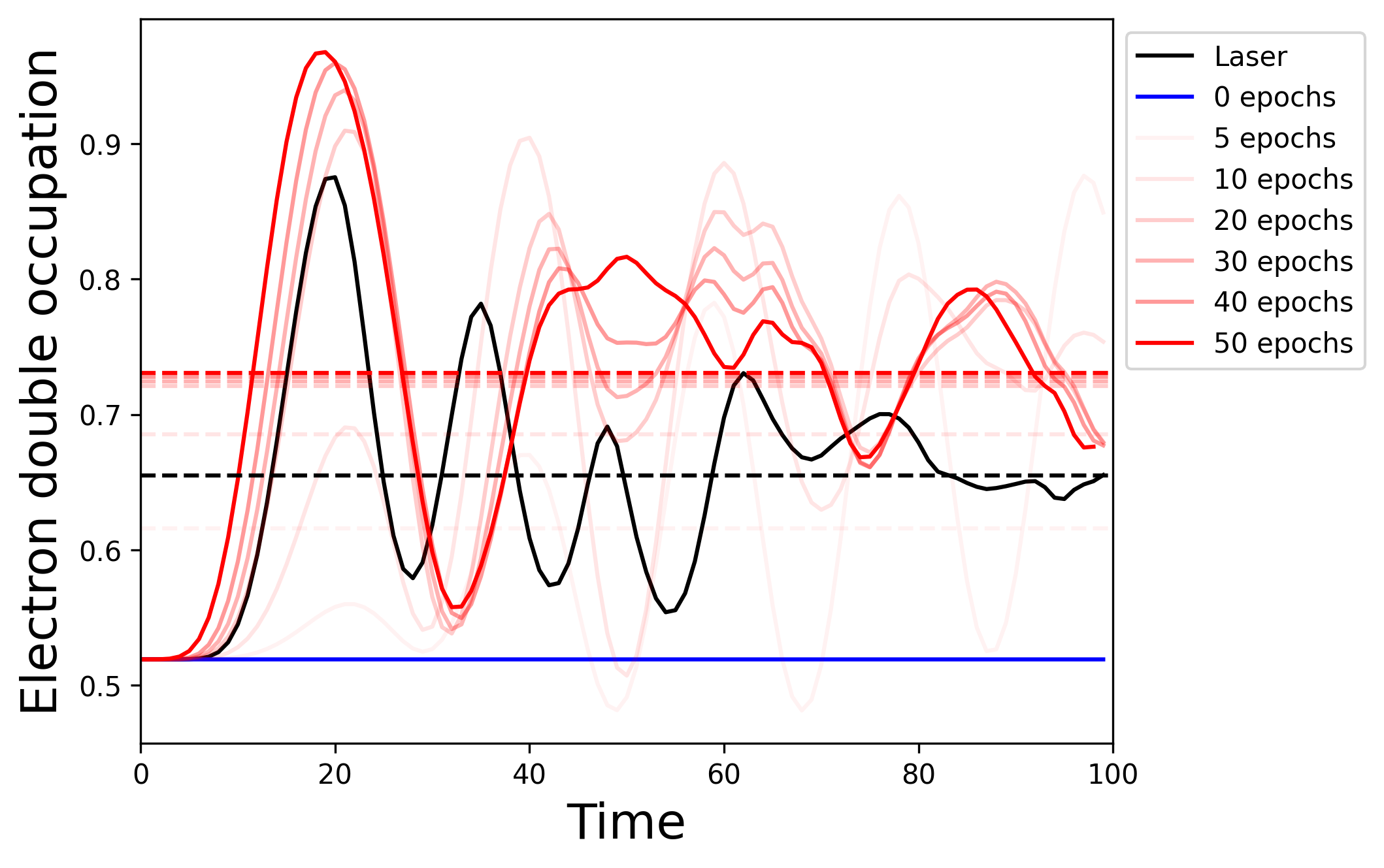}
\caption{Development of the feed-forward controller during training. Solid curves show the total double occupation at the indicated training epochs and for the monochromatic reference. Dashed lines indicate the corresponding means over the displayed segments. The mean at initialization is 0.519; after 50 epochs it is 0.731, compared with 0.655 for the reference. The averaging interval differs from that of the shorter FNN segment in Fig.~\ref{Result1}.}
\label{Training}
\end{figure}

The separate FNN training sequence in Fig.~\ref{Training} illustrates how the control develops during optimization. The reported mean grows from 0.519 at initialization to 0.731 after 50 training epochs, above the 0.655 reference for that displayed comparison. This comparison is distinct from the shorter FNN segment in Fig.~\ref{Result1}, and the two averages refer to different displayed intervals. The progression shows that even an instantaneous observable-to-field map can learn a useful nonlinear driving strategy.

\subsection{Predicting the driven observables}

\begin{figure}[t]
\centering
\includegraphics[width=0.91\columnwidth]{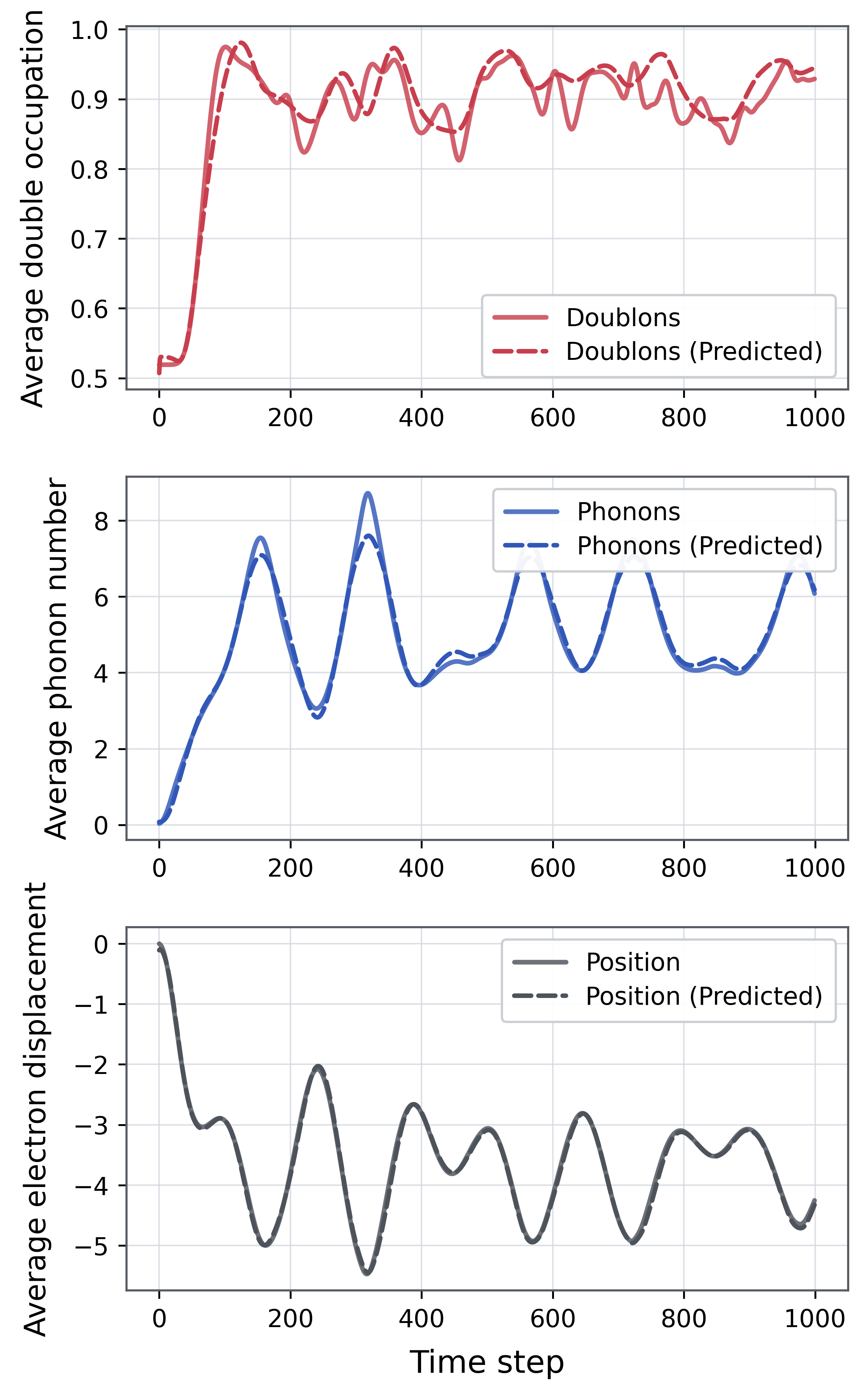}
\caption{Simulator observables (solid) and recurrent-network predictions (dashed) over 1000 one-step evaluations. From top to bottom: total double occupation $D$, local phonon number $n$, and local phonon coordinate $X$. The inputs at every step are the preceding simulator observables and the applied field. The comparison therefore evaluates teacher-forced one-step prediction, rather than a free-running 1000-step rollout. Parameters are those of Sec.~II; training episodes contain 200 steps with $\delta t=0.1$ and learning rate $5\times10^{-6}$. The upper ordinate is interpreted as total double occupation according to Eq.~\eqref{doublon}; the lower ordinate denotes the phonon coordinate.}
\label{Predictions}
\end{figure}

Figure~\ref{Predictions} compares the recurrent predictor with the simulator over 1000 one-step evaluations. The network follows the main temporal structure of all three observables, including the initial rise in double occupation and the oscillatory phonon response. Table~\ref{tab:prediction} reports the mean absolute error (MAE) and root-mean-square error (RMSE) for this sequence. For an observable $y$ with $K$ evaluated samples, these are
\begin{align}
\mathrm{MAE}_y&=\frac{1}{K}\sum_{k=1}^{K}|\widetilde y_k-y_k|,\nonumber\\
\mathrm{RMSE}_y&=\left[\frac{1}{K}\sum_{k=1}^{K}(\widetilde y_k-y_k)^2\right]^{1/2}.
\end{align}

\begin{table}[t]
\caption{Reported one-step prediction errors on the sequence in Fig.~\ref{Predictions}. The final column divides MAE by the observed range of the corresponding simulator trace (0.456, 8.680, and 5.470, respectively). It is a descriptive normalization, not a confidence interval or a pointwise relative error.}
\label{tab:prediction}
\begin{ruledtabular}
\begin{tabular}{lccc}
Observable & MAE & RMSE & MAE/range\\
\hline
Total double occupation & 0.0222 & 0.0287 & $4.9\%$\\
Local phonon number & 0.163 & 0.234 & $1.9\%$\\
Local phonon coordinate & 0.0375 & 0.0493 & $0.69\%$\\
\end{tabular}
\end{ruledtabular}
\end{table}

On this range-normalized measure, the phonon coordinate and phonon number are reproduced particularly closely. Double occupation has the largest range-normalized error despite its larger weight in the loss, leaving the greatest relative discrepancy for this observable in the evaluated sequence. The evaluation sequence is longer than a training episode, but each prediction is conditioned on simulator inputs.

\section{Conclusion and outlook}

In this work, we have presented a differentiable control approach of a finite two-site nonlinear electron-phonon model with quadratic coupling. We compared different methods, namely selected vector, B-spline, and recurrent-network controls, that increase the time-averaged total double occupation relative to a monochromatic reference, typically used so far in these tasks. A recurrent dynamics network also reproduces the next-step observables with small mean absolute errors on one recorded trajectory. These results establish a computational proof of concept for control and prediction in this model, suggesting directions to improve control, induced and sustained superconductivity or other non-equilibrium phases of matter. Overall, this work opens the door towards the optimization of control of complex systems, with the aim to manipulate the phases of matter of real materials.\\

\textit{AI-assisted research tools} — OpenAI ChatGPT (model gpt-6-Astra) assisted with manuscript drafting and revision. The authors retain responsibility for the scientific content and conclusions.

\bibliographystyle{apsrev4-2}
\bibliography{bibliography}

\end{document}